\documentclass[preprint,sort&compress,numbers]{elsarticle} % choose Elsevier style article as document time

\usepackage{amsmath}
\usepackage{amssymb}
\usepackage{subfigure}
\usepackage{caption}

\usepackage{xcolor}
\newcommand{\ignore}[1]{}

\newcommand{\bbm}{\begin{bmatrix}}
\newcommand{\ebm}{\end{bmatrix}}
\newcommand{\bma}[1]{\left[\begin{array}{#1}}
\newcommand{\ema}{\end{array}\right]}

\DeclareMathAlphabet{\mbf}{OT1}{ptm}{b}{n}
\newcommand{\mbs}[1]{{\boldsymbol{#1}}}
\def\dotb{{\raisebox{-0.6ex}{ \kern0.2ex\raisebox{0.8ex}{\tiny $\circ$}}}}
\def\ddota{{\raisebox{-0.6ex}{ \raise0.2ex\hbox{ \LARGE $\cdot\hspace*{-0.2ex}\cdot$}}}}
\def\ddotb{{\raisebox{-0.6ex}{ \kern0.2ex\raisebox{0.8ex}{\tiny $\circ\circ$}}}}

\newcommand{\trans}{{\ensuremath{\mathsf{T}}}} % transpose
\newcommand{\beq}{\begin{equation}}
\newcommand{\eeq}{\end{equation}}
\newcommand{\bdis}{\begin{displaymath}}
\newcommand{\edis}{\end{displaymath}}
\newcommand{\beqarray}{\begin{eqnarray}}
\newcommand{\eeqarray}{\end{eqnarray}}
\newcommand{\beqarraynn}{\begin{eqnarray*}}
\newcommand{\eeqarraynn}{\end{eqnarray*}}

\journal{Acta Astronautica} % specify journal
\begin{document}

%%%%%%%%%%%%%%%%%%%%%%%%% %%%%%%%%%%%%%%%%%%%%%%%%%%%%%%%%%%%%%%%%%%%%%%%%%%%%%%
%													     FRONTMATTER 
%%%%%%%%%%%%%%%%%%%%%%%%% %%%%%%%%%%%%%%%%%%%%%%%%%%%%%%%%%%%%%%%%%%%%%%%%%%%%%%
\begin{frontmatter}
% Article Title
\title{Passivity-Based Robust Shape Control of a Cable-Driven Solar Sail Boom for the CABLESSail Concept}

% Author information 
\author[label1]{Soojeong Lee}%\fnref{fn1}}
%\ead{lee03939@umn.edu}
\author[label1]{Ryan J. Caverly}% \fnref{fn2}}
%\ead{rcaverly@umn.edu}

\affiliation[label1]{organization={Department of Aerospace Engineering \& Mechanics, University of Minnesota},
            city={Minneapolis}, 
            state={MN},
            postcode={55455},
            country={United States}}

\begin{abstract}
    Solar sails provide a means of propulsion using solar radiation pressure, which offers the possibility of exciting new spacecraft capabilities. However, solar sails have attitude control challenges because of the significant disturbance torques that they encounter due to imperfections in the sail and its supporting structure, as well as limited actuation capabilities.
    The Cable-Actuated Bio-inspired Lightweight Elastic Solar Sail (CABLESSail) concept was previously proposed to overcome these challenges by controlling the shape of the sail through cable actuation. The structural flexibility of CABLESSail introduces control challenges, which necessitate the design of a robust feedback controller for this system.
    The purpose of the proposed research here is to design a robust controller to ensure precise and reliable control of CABLESSail's boom. Taking into account the system dynamics and the dynamic properties of the CABLESSail concept, a passivity-based proportional-derivative (PD) controller for a single boom on the CABLESSail system is designed. To reach the nonzero desired setpoints, a feedforward input is additionally applied to the control law and a time-varying feedforward input is used instead of the constant one to effectively track a time-varying desired boom tip deflection. 
    This control law is assessed by numerical simulations and by tests using a smaller-scale prototype of Solar Cruiser.
    Both the simulation and the test results show that this PD control with the time-varying feedforward input robustly controls the flexible cable-actuated solar sail.
\end{abstract}

\begin{keyword}
Passivity-Based Control \sep Solar Sails \sep Robust Control \sep Flexible Systems \sep Cable Actuation \sep Feedforward Control
\end{keyword}

\end{frontmatter}

%%%%%%%%%%%%%%%%%%%%%%%%%%%%%%%%%%%%%%%%%%%%%%%%%%%%%%%%%%%%%%%%%%%%%%%%%%%%%%%
%													   INTRODUCTION
%%%%%%%%%%%%%%%%%%%%%%%%%%%%%%%%%%%%%%%%%%%%%%%%%%%%%%%%%%%%%%%%%%%%%%%%%%%%%%%

\section{Introduction}

A solar sail is a non-traditional spacecraft that uses solar radiation pressure for propulsion rather than the typical use of fuel. This makes it a promising concept for futuristic space missions including interstellar travel~\cite{gong2019review,zeng2012optimal,berthet2024space,miller2022high} and achieving orbits that traditionally require tremendous amounts of fuel, such as orbits outside of the ecliptic plane~\cite{thomas2020solar,kobayashi2020high,pezent2021SolarCruiser} and reaching targets beyond low-Earth orbit~\cite{ANCONA2019601,LANTOINE202477,KARLAPP2024889}. However, solar radiation pressure is so weak that very large and lightweight solar sails are desired in order to maximize their efficiency and effectiveness. This results in solar sails with significant structural flexibility and large disturbance torques due to imperfections in the sail (e.g., wrinkling or billowing)~\cite{gong2019review} and the spacecraft structure (e.g., thermal expansion/contraction or vibrations)~\cite{gauvain2023solar,brownell2023time,Boni2023-mb,ingrassia2013solar,JIN201473}, which necessitates an emphasis on the solar sail's attitude control and momentum management system~\cite{spencer2019solar,inness2023momentum}. There have been traditional and novel attitude control actuators proposed for use with solar sails, for example, using reaction wheels~\cite{polites2008solar,gong2021attitude}, mounting sliding masses~\cite{wie2007solar,bolle2008solar,lappas2009practical,romagnoli2011high,adeli2011scalable,steyn2011cubesat,huang2019solar}, variable reflectivity control devices~\cite{funase2012modeling,borggrafe2014optical,davoyan2021photonic,ullery2018strong}, and variable-shape mechanisms~\cite{Chujo2022-oq}. Although these actuators are being considered in the design of existing solar sails (e.g., Solar Cruiser~\cite{pezent2021SolarCruiser,inness2023momentum, inness2024controls}), they all have limitations in control authority or scalability that will be exacerbated as solar sails are designed with larger sizes and travel to further distances from Earth.
Another attitude control method based on the shape variation of booms using piezoelectric actuator has been introduced to overcome these limitations~\cite{zhang2021solar,zhang2021three}, but it has limitations in magnitude of torque generated and long-term operation due to continuously required voltage usage for keeping the piezoelectric actuator on.

\begin{figure}[!t]%[htb]
	\centering\includegraphics[width=4.0in]{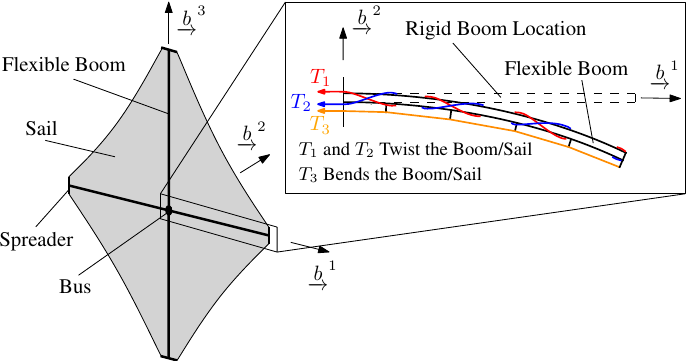}%
	\caption{An illustration of the CABLESSail concept, where actuating cables are routed down the length of the solar sail's structural booms.}
	\label{fig:CABLESSail}
\end{figure}

The Cable-Actuated Bio-inspired Lightweight Elastic Solar Sail (CABLESSail) concept~\cite{CaverlyISSS}, shown in Figure~\ref{fig:CABLESSail}, was recently proposed as a new solar sail actuation technology. 
The CABLESSail concept involves the use of cables routed down the length of the solar sail's booms that are tensioned appropriately to purposefully deflect the booms and induce a change in the shape of the sail. This shape change induces an imbalance in solar radiation pressure, resulting in the generation of control torques. Specifically, bending of the booms results in the generation of pitch and yaw torques. Bending the booms can also result in a roll torque by creating asymmetric shapes similar to those described in~\cite{gauvain2023solar} in the presence of a non-zero Sun incidence angle. Also, in contrast to the piezoelectric boom deformation approach described in~\cite{zhang2021solar,zhang2021three}, a motor brake can be used to hold tensions in the actuating cable without any power draw, allowing for the boom to be held in a deformed position for long durations. 
Preliminary analysis of the CABLESSail concept~\cite{CaverlyISSS,BunkerSciTech} has shown that if the booms can be deformed as desired, it should be capable of generating large control torques in all three axes of a solar sail, which will allow for the reliable attitude control and the mitigation of disturbance torques.
Also, the lightweight and stowable nature of CABLESSail's actuating cables will facilitate its use on large, next-generation solar sails with large inertia.
Nevertheless, the CABLESSail concept has several control challenges that need to be addressed before the technology is matured. 
In particular, a robust feedback controller is to be designed that is capable of reliably controlling the deformation of the solar sail's booms using cable actuation. As shown in~\cite{BunkerSciTech}, open-loop actuation of the cables results in undesirable long-term vibration of the entire solar sail structure, which necessitates the use of feedback control. Part of the challenge is that tremendous uncertainty exists in the structural dynamics models of solar sail booms, which must be explicitly accounted for in the design of CABLESSail's feedback control system~\cite{BunkerSciTech,lee2024robust}.

Passivity-based control is a robust control technique that has successfully been applied to a variety of systems, including flexible space structures and robots~\cite{hassanpour2018linear,hassanpour2020collocated}, as well as cable-actuated robotic systems~\cite{Caverly2014Passivity,Caverly2014Noncollocated,Caverly2018Manipulator,Godbole2019}. The main benefit of passivity-based control is that rather than relying on exact knowledge of a system model to guarantee closed-loop stability, the only knowledge required is whether the system being controlled is passive~\cite{MarquezNonlinear}. Many mechanical systems feature passive input-output mappings that are robust to large parameter variations, which allows for passivity-based control to be applied to these systems with robust guarantees.

This paper focuses on the use of passivity-based control as a means to robustly control the bending deformation of a single CABLESSail boom using cable actuation. The main contributions of this paper include the maturation of the CABLESSail concept through 1) the derivation of a mathematically-rigorous robust controller that allows for the reliable deformation of a solar sail's booms and 2) the validation of the proposed controller through simulation and experimental implementation. A contribution specific to this work in comparison to the preliminary results presented in~\cite{lee2024robust} is the experimental validation of the proposed control method on a small-scale prototype testbed. Beyond the scope of CABLESSail, the robust control formulation developed in this paper provides insight into how cable actuation may be implemented in the control of other large flexible space structures.

The remainder of this paper proceeds as follows. A dynamic model of a CABLESSail boom is introduced, which focuses on a single CABLESSail boom actuated with a single cable. This model is used to compute the equilibrium boom deflections associated with varying amounts of cable tension, as well as linearized system dynamics about these equilibrium points. Then, the system's linearized dynamics are used to analyze the system's passivity properties.
Then, a robust proportional-derivative (PD) controller is formulated with a time-varying feedforward input using passivity-based control theory.
Simulation results are included that demonstrate the closed-loop performance of the proposed passivity-based control approach when tracking constant and time-varying desired boom tip displacements. Furthermore, test results using a prototype are provided with discussion on mapping between a torque and a tip deflection. Overall discussions both on simulation and test results follow them. Finally, conclusions and future works are presented.

%%%%%%%%%%%%%%%%%%%%%%%%%%%%%%%%%%%%%%%%%%%%%%%%%%%%%%%%%%%%%%%%%%%%%%%%%%%%%%%%%%%%%%%%%%%%%%%%%%%%%
\section{Dynamic System} \label{sec:dynamic}

In this section, the dynamic model of a single CABLESSail boom is introduced for use in control design and analysis. Furthermore, the desired setpoint of the controller is calculated by solving for the system’s forced equilibrium points and then a linearization of the system is derived for use in the design of the controller presented in the next section.

\subsection{Dynamic Equations of a Single CABLESSail Boom}
\begin{figure}[!t]%[htb]
	\centering\includegraphics[width=0.99\textwidth]{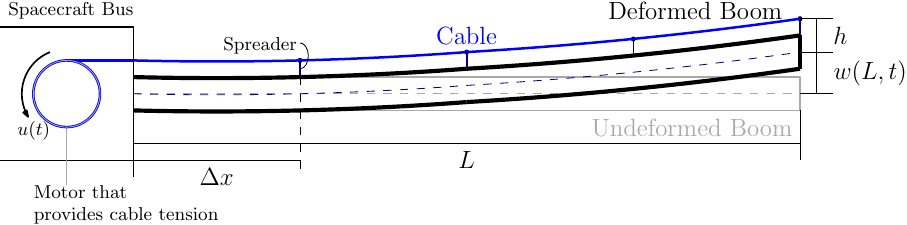}%[width=4.0in]
	\caption{A single boom with a single cable of CABLESSail.}
	\label{fig:SingleBoom}
\end{figure}
The equations of motion of a single CABLESSail boom actuated with a single cable, as shown in Figure~\ref{fig:SingleBoom}, were derived in~\cite{BunkerSciTech} using an assumed-modes method~\cite{liu2014dynamics} and an Euler-Bernoulli beam model. In this formulation, the transverse deformation of the boom at spatial location $x$ along the length of the boom at time $t$ is given by $w(x,t) = \mbs{\Psi}(x)\mbf{q}(t)$, where $\mbf{q}(t) \in \mathbb{R}^n$ contains the coefficients of the assumed-mode basis functions and $\mbs{\Psi}(x) \in \mathbb{R}^{1 \times n}$ contains the assumed-mode basis functions. Here, the dimension $n$ is determined by the number of basis functions used to represent the deformations.
The kinetic energy and potential energy of the system are respectively defined as
\begin{equation}
	\label{eq:dyn_kinE}
	T(t) = \frac{1}{2}\dot{\mbf{q}}^\trans(t) \mbf{M}\dot{\mbf{q}}(t),
\end{equation}
\begin{equation}
	\label{eq:dyn_potE}
	V(t) = \frac{1}{2}\mbf{q}^\trans(t) \mbf{K}\mbf{q}(t),
\end{equation}
where $\mbf{M} = \mbf{M}^\trans \in \mathbb{R}^{n \times n}$ is the system's mass matrix, $\mbf{K} = \mbf{K}^\trans \in \mathbb{R}^{n \times n}$ is the system's stiffness matrix. These matrices are mathematically defined as
\begin{equation}
	\label{eq:nonlinearAdd1}
	\mbf{M} = \int_{0}^{L} \rho \mbs{\Psi}^\trans(x) \mbs{\Psi}(x) \mathrm{d}x,
\end{equation}
\begin{equation}
	\label{eq:nonlinearAdd2}
	\mbf{K} = EI \int_{0}^{L} \left(\frac{\partial^2 \mbs{\Psi}}{\partial x^2}\right)^\trans  \left(\frac{\partial^2 \mbs{\Psi}}{\partial x^2}\right)  \mathrm{d}x,
\end{equation}
where $L$ is the boom length, $\rho$ the linear density of the boom, $E$ the modulus of elasticity, and $I$ the second moment of area. Using a Lagrangian approach, the equations of motion are derived in~\cite{BunkerSciTech} and are given by
\begin{equation}
	\label{eq:nonlinear}
	\mbf{M} \ddot{\mbf{q}}(t) +\mbf{K}\mbf{q}(t) = \mbf{f}(\mbf{q}(t),u(t)),
\end{equation}
where $u(t) \in \mathbb{R}$ is the control input tension applied to the cable and $\mbf{f}(\mbf{q}(t),u(t)) \in \mathbb{R}^n$ contains the actuation forces acting on the system described by
\begin{equation}
\label{eq:nonlinearAdd3}
\mbf{f}(\mbf{q}(t),u(t)) = \frac{\mbs{\Psi}^\ast}{\Delta x}\mbf{q}(t)u(t)
 + h\left(\frac{\partial \mbs{\Psi}}{\partial x}\right)^\trans_{x=L}u(t),
\end{equation}
where $h$ is the distance between the cable attachment points and the boom's neutral axis, $\Delta x$ is the distance between cable attachment points, and $\mbs{\Psi}^\ast$ is a constant matrix with respect to time that depends on the number of spreaders used, $n_s$, and computes the reaction forces perpendicular to the deflected boom due to the tension within the cable. A fixed boundary condition is assumed at the root of the boom, which requires the spatial basis functions to satisfy $\mbs{\Psi}(0) = \mbf{0}$ and $\partial \mbs{\Psi}/\partial x|_{x = 0} = \mbf{0}$. Following the work of~\cite{BunkerSciTech}, the assumed-mode basis functions are chosen as $\mbs{\Psi}(x)=\bbm x^2&&x^3&&x^4\ebm$, which was found to capture the first couple of vibrational modes well with relatively few basis functions.

The ODE in~\eqref{eq:nonlinear} is converted to first-order state-space form by defining the state $\mbf{x}(t) = \bbm \mbf{q}(t)^\trans & \dot{\mbf{q}}(t)^\trans \ebm^\trans$. This results in
\begin{equation}
\label{eq:FirstOrder}
\dot{\mbf{x}}(t)
= \bbm \dot{\mbf{q}}(t) \\ \ddot{\mbf{q}}(t) \ebm
= \bbm \dot{\mbf{q}}(t) \\ \mbf{M}^{-1}\left(\frac{\mbs{\Psi}^\ast}{\Delta x}\mbf{q}(t)u(t)
 + h\left(\frac{\partial \mbs{\Psi}}{\partial x}\right)^\trans_{x=L}u(t)
 - \mbf{K}\mbf{q}(t)\right) \ebm.
\end{equation}

The numerical parameters used in the simulation-based tests are chosen based on the properties of Solar Cruiser's triangular rollable and collapsible (TRAC) booms~\cite{JeremyTRAC, nguyen2023solar}. These numerical values are listed in Table~\ref{table:SimParam} and used for all simulation studies in Sections~\ref{subsec:Eqpt} to~\ref{sec:simulation}.

\begin{table}
\centering
\begin{tabular}{cccc}
    \hline
    Symbol & Parameter & Value & Units \\
     \hline
    $L$ & Boom Length & $29.4$ & m \\
    $\rho$ & Linear Density & $0.1$ & kg/m \\
    $E$ & Modulus of Elasticity & $228$ & GPa\\
    $I$ & Second Moment of Area & $4.99\times10^{-10}$ &m$^4$ \\
    $h$ & Distance Between the Cable and the Beam & $0.1$ & m \\
    $n_s$ & Number of Spreaders & $10$ & - \\
    $\Delta x$ & Distance Between Nodes & $2.94$ & m \\
    \hline
\end{tabular}
\caption{Numerical values for the simulation-based tests throughout the paper.}
\label{table:SimParam}
\end{table}

\subsection{Solving for the System's Forced Equilibrium Points } \label{subsec:Eqpt}
\begin{figure}[!t]%[htb]
	\centering\includegraphics[width=3.5in]{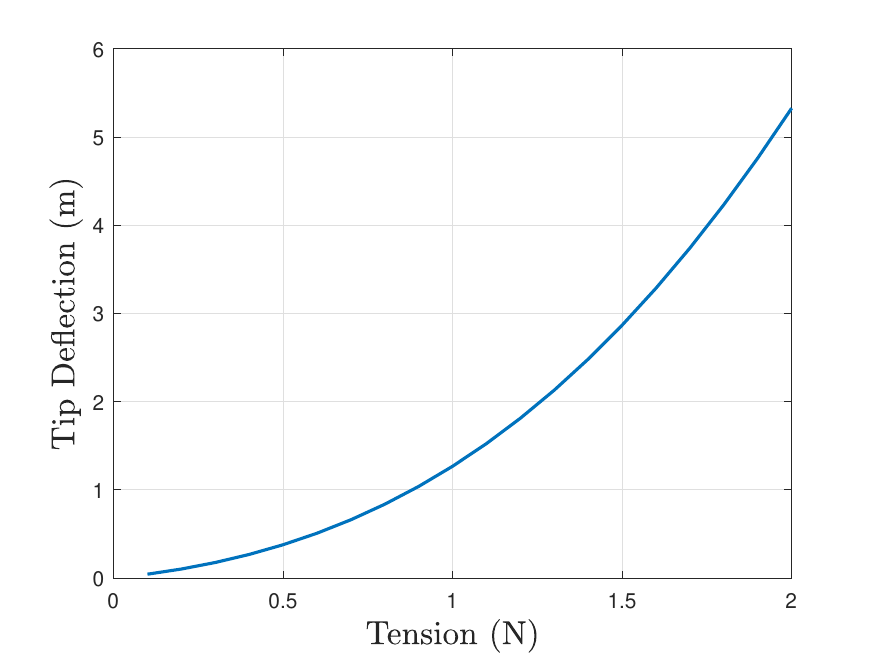}
	\caption{Equilibrium boom tip deflection as a function of equilibrium tension.}
	\label{fig:desired}
\end{figure}

The equilibrium point of the system depends on the amount of tension applied to the actuating cable.
For a fixed value of tension, $\bar{u}=T_{eq}$,~\eqref{eq:FirstOrder} is used to solve for the forced equilibrium state $\mbf{q}_{eq}$ that results in $\dot{\mbf{x}}(t) = \mbf{0}$, given by
\begin{equation}
\label{eq:EqPt}
\mbf{q}_{eq} = \bar{\mbf{q}}(T_{eq}) = \left(\mbf{K}-\frac{\mbs{\Psi}^\ast}{\Delta x}T_{eq}\right)^{-1}h\left(\frac{\partial \mbs{\Psi}}{\partial x}\right)^\trans_{x=L} T_{eq}.
\end{equation}
A plot of the equilibrium tip deflection $w_{eq}(L) = \mbs{\Psi}(L)\mbf{q}_{eq}$ is provided in Figure~\ref{fig:desired} based on the parameters in Table~\ref{table:SimParam}, where it is seen that the relationship between equilibrium deflection on equilibrium tension is quadratic.

The control formulation presented in this work makes use of the forced equilibrium point as a desired setpoint $\mbf{q}_{des}=\mbf{q}_{eq}$. 
Therefore, the plot in Figure~\ref{fig:desired} can be used to determine the equilibrium tension, $T_{eq}$, required to maintain a desired tip deflection setpoint in equilibrium. Once the value of $T_{eq}$ is chosen, the desired state $\mbf{q}_{des}$ can be obtained using~\eqref{eq:EqPt}.
It can be shown numerically that in the given range of $T_{eq}$ from $0$~N to $2$~N, the term $\left(\mbf{K}-\frac{\mbs{\Psi}^\ast}{\Delta x}T_{eq}\right)$ that appears in~\eqref{eq:EqPt} is invertible.

\subsection{Linearization at the Equilibrium Point}
To assist with the passivity analysis performed in Section~\ref{sec:PassivityAnalysis}, it is convenient to derive a linearization of the system about the equilibrium point $\bar{\mbf{x}}(T_{eq})$ and $\bar{u}(T_{eq})$. The system state and input are described based on perturbations from the equilibrium point as
\begin{equation}
\label{eq:LinX}
\mbf{x}(t)=\bar{\mbf{x}}(T_{eq})+\delta \mbf{x}(t),
\end{equation}
\begin{equation}
\label{eq:LinU}
u(t)=\bar{u}(T_{eq})+\delta u(t),
\end{equation}
where the equilibrium points are rewritten from~\eqref{eq:EqPt} in the form
\begin{equation}
\label{eq:EqPtX}
\bar{\mbf{x}}(T_{eq})
=\bbm
\bar{\mbf{q}}(T_{eq})\\
\dot{\bar{\mbf{q}}}
\ebm
=\bbm
\left(\mbf{K}-\frac{\mbs{\Psi}^\ast}{\Delta x}T_{eq}\right)^{-1}h\left(\frac{\partial \mbs{\Psi}}{\partial x}\right)^\trans_{x=L} T_{eq}\\
\mbf{0}
\ebm,
\end{equation}
\begin{equation}
\label{eq:EqPtU}
\bar{u}(T_{eq})=T_{eq}.
\end{equation}
The linearized equations of motion are written in state-space form as
\begin{equation}
\label{eq:LinEoM}
\frac{\mathrm{d}(\delta \mbf{x})}{\mathrm{d}t}=\mbf{A}(T_{eq})\delta \mbf{x}+ \mbf{B}(T_{eq}) \delta u,
\end{equation}
where
\begin{align}
\mbf{A}(T_{eq}) &= 
\bbm
\mbf{0} & \mbf{1}\\
\mbf{M}^{-1}\left(\frac{\mbs{\Psi}^\ast}{\Delta x}T_{eq}-\mbf{K}\right) & \mbf{0}
\ebm, \\
\mbf{B}(T_{eq}) &= 
\bbm
\mbf{0}\\
\mbf{M}^{-1}\left(\frac{\mbs{\Psi}^\ast}{\Delta x}\left(\mbf{K}-\frac{\mbs{\Psi}^\ast}{\Delta x}T_{eq}\right)^{-1}h\left(\frac{\partial \mbs{\Psi}}{\partial x}\right)^\trans_{x=L} T_{eq} + h\left(\frac{\partial \mbs{\Psi}}{\partial x}\right)^\trans_{x=L}\right)
\ebm.
\end{align}

\section{Control Laws}
In this section, a passivity-based control approach is introduced and applied to the dynamic model from Section~\ref{sec:dynamic}. To ensure stability of the closed-loop system with the proposed passivity-based controller, the passivity of the linearized system is analyzed. Then, a PD controller with a time-varying feedforward input is formulated to track a time-varying desired boom tip deflection.

\subsection{Overview of Passivity Theory}

Passivity-based control can ensure the stability of the closed-loop system when the open-loop system is proved to be passive. This is determined by the passivity theorem, which states that the negative feedback interconnection of a passive system and a very strictly passive system is input-output stable, which implies the closed-loop is asymptotically stable if both systems have minimal realizations~\cite{MarquezNonlinear}. A single-input single-output (SISO) linear time-invariant (LTI) system is known to be passive if its phase lies within $[-90^\circ,+90^\circ]$ and very strictly passive if its phase lies within $(-90^\circ,+90^\circ)$ and has strictly positive feedthrough. The implication of this theorem for passivity-based control is that any passive system can be controlled by a very strictly passive negative feedback controller.

\subsection{Passivity Analysis}
\label{sec:PassivityAnalysis}
The system output considered in this paper is the transverse deflection rate of the tip, which can be expressed as
\begin{equation}
\label{eq:Output}
y(t) = \dot{w}(L,t) = \mbs{\Psi}(L)\dot{\mbf{q}}(t).
\end{equation}

In order to make use of passivity-based control, it must be verified that the mapping from input $u(t)$ to output $y(t)$ is in fact passive. To perform this verification, the passivity properties of the system's linearization about different equilibrium points are assessed. Although this process is not capable of providing a guarantee that the nonlinear system is passive, it is a practical preliminary assessment of passivity.

To proceed with this analysis, the output equation in~\eqref{eq:Output} is rewritten as
\begin{equation}
\label{eq:LinOutput}
\delta y=\mbf{C}\delta \mbf{x}+ D \delta u % \mathrm{}
=\bbm
\mbf{0} & \mbs{\Psi}(L)
\ebm
\delta \mbf{x}+
\bbm
0
\ebm
\delta u.
\end{equation}

\begin{figure}[!t]%[htb]
	\centering
	\subfigure[]
	{
		\includegraphics[width=0.7\textwidth]{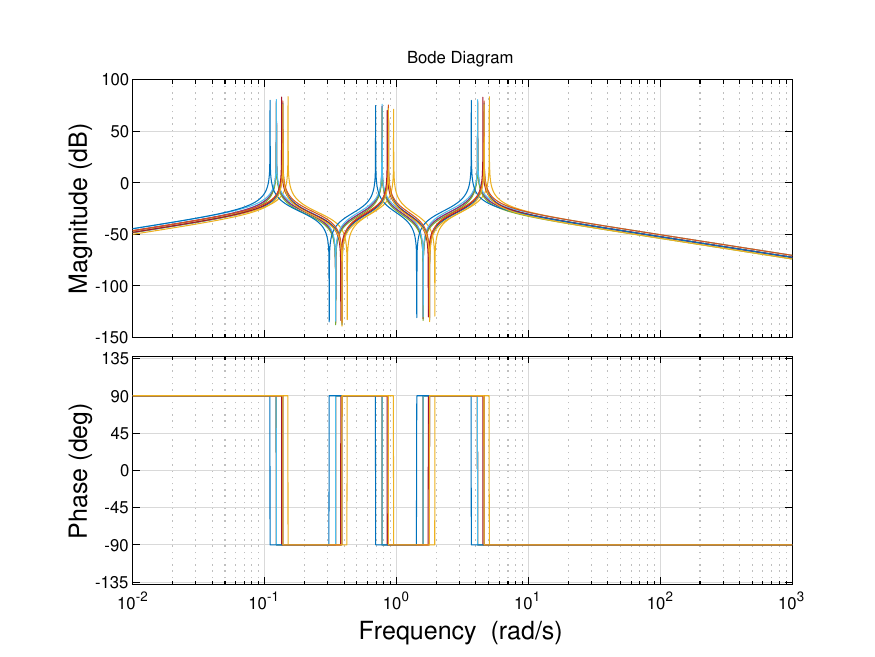}
	}
	\subfigure[]
	{%[width=3.5in]
		\includegraphics[width=0.7\textwidth]{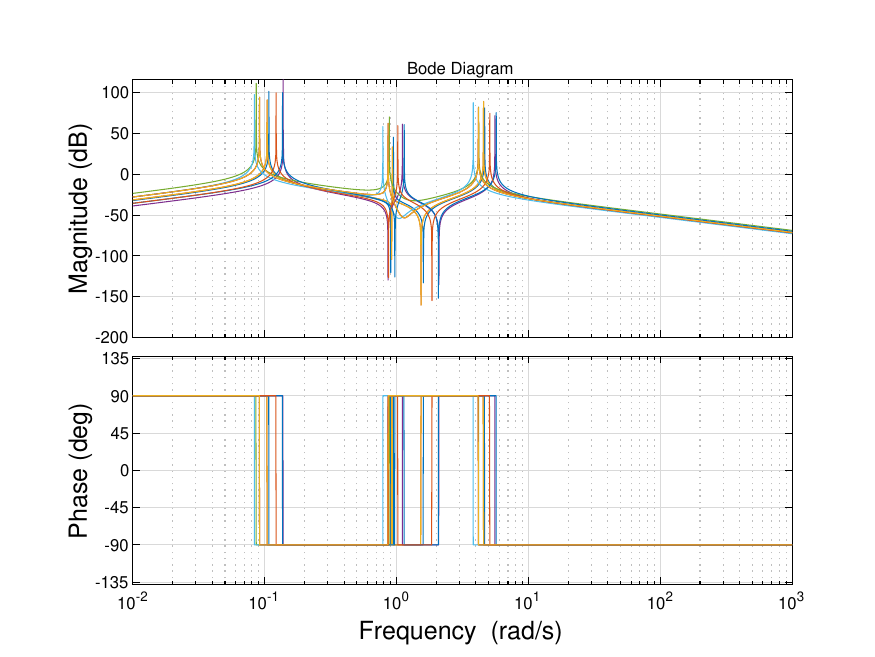}
	}
	\caption{Bode plots of the uncertain linearized system remaining passive with the transverse deflection rate of the boom tip as an output when (a) $T_{eq}=0$~N and (b) $T_{eq}=1$~N.}
	\label{fig:Bode}
\end{figure}

\begin{figure}[!t]%[htb]
	\centering
	\subfigure[]
	{
		\includegraphics[width=0.7\textwidth]{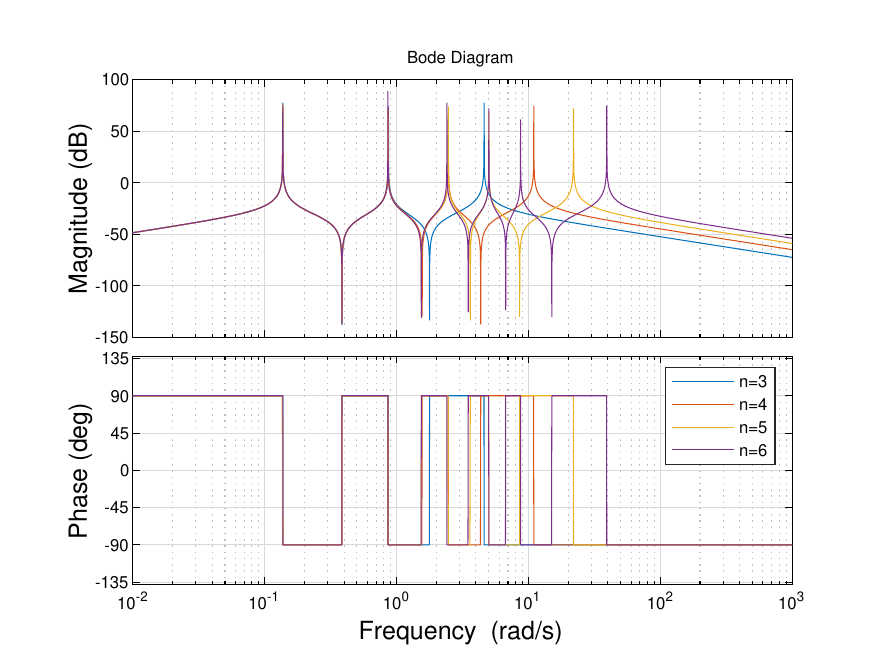}
	}
	\subfigure[]
	{%[width=3.5in]
		\includegraphics[width=0.7\textwidth]{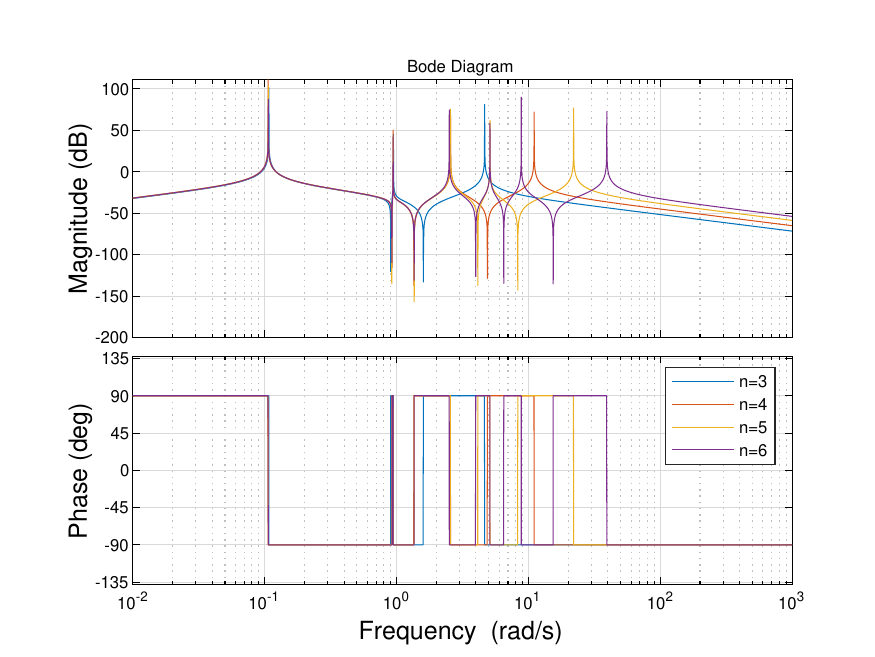}
	}
	\caption{Bode plots of the linearized system with a varying number of assumed modes, $n$, remaining passive with the transverse deflection rate of the boom tip as an output when (a) $T_{eq}=0$~N and (b) $T_{eq}=1$~N.}
	\label{fig:Bode_assumed_modes}
\end{figure}

Bode plots of the linearized system defined by~\eqref{eq:LinEoM} and~\eqref{eq:LinOutput} for the equilibrium points associated with $T_{eq} = 0$~N and $T_{eq} = 1$~N are included in Figure~\ref{fig:Bode} and Figure~\ref{fig:Bode_assumed_modes}. The Bode plots in Figure~\ref{fig:Bode} include uncertainty in the range of $\pm20$\% added to the nominal values of elastic modulus, linear density, and second moment of area provided in Table~\ref{table:SimParam}, which directly relate to uncertainty in the boom's mass and stiffness properties. The Bode plots in Figure~\ref{fig:Bode_assumed_modes} include models with a varying number of assumed modes with the nominal parameter values. Higher-order polynomial basis functions are used for the additional basis functions. These plots show that the phase of the system's linearization always remains between $\pm90^\circ$ for the entire uncertainty range and for all models regardless of the number of assumed modes used, which indicates that the linearized system is robustly passive.

\subsection{Proportional-Derivative Controller}
Knowing that the system's linearization is passive%~\cite{lee2024robust}
, any very strictly passive controller can be designed to guarantee closed-loop stability. In particular, a proportional-derivative (PD) controller is chosen in this paper, which uses proportional (P) and derivative (D) feedback to control the dynamic system.

\begin{figure}[!t]%[htb]
	\centering\includegraphics[width=0.99\textwidth]{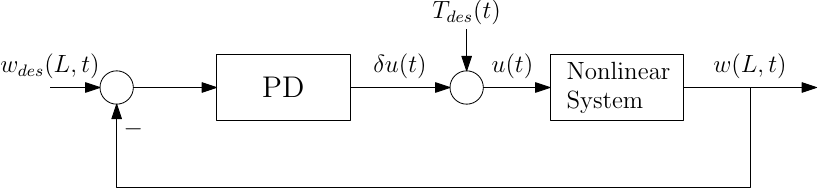}%[width=4.0in]
	\caption{Block diagram of the closed-loop system with time-varying feedforward PD control.}
	\label{fig:PIDNonBlockVarying}
\end{figure}

The control input $u(t)$ is composed of a feedback control input $\delta u(t)$ and a feedforward control input $T_{des}(t)$,
\begin{equation}
\label{eq:ControlInput_u}
u(t) = T_{des}(t) + \delta u(t),
\end{equation}
as expressed in Figure~\ref{fig:PIDNonBlockVarying}.
The feedforward input helps the system achieve the non-zero desired setpoint and the feedback input makes up the error of the estimates or the uncertainties in the real system.

The feedback control input $\delta u(t)$ is given by
\begin{equation}
\label{eq:feedbackInput}
\delta u(t) = -k_p(w(L,t) - w_{des}(L)) - k_d(\dot{w}(L,t) - \dot{w}_{des}(L)),
\end{equation}
where $w(L,t)$ and $\dot{w}(L,t)$ are the responses from the system output, and $w_{des}(L)$ and $\dot{w}_{des}(L)$ are the desired setpoints for the controller. Since the system output is the velocity of the tip, $y(t) = \dot{w}(L,t)$, the proposed PD controller acting on the boom tip transverse deflection can be thought of as a proportional-integral (PI) controller on the output $y(t)$, which is the boom tip transverse deflection rate. A PI controller is very strictly passive (i.e., its phase is within $(-90^\circ,+90^\circ)$ and it has strictly positive feedthrough $k_d > 0$), therefore, the passivity theorem guarantees that the closed-loop system is robustly input-output stable with this chosen controller. The significance of this result lies in the guarantee of robust input-output stability of the closed-loop system in the presence of significant variations in the structural properties of the cable-actuated boom structure.

The feedforward control input $T_{des}(t)$ is related to the equilibrium tension of the system. To avoid large sudden changes in the system's control inputs, it is desirable for the desired reference tip deflection and its associated equilibrium feedforward control input to smoothly vary in time. This motivates the design of a feedforward control input $T_{des}(t)$ based on a maneuver that nominally takes the system from an initial deflection to a final desired deflection. To achieve this, the time-varying feedforward input is chosen to have zero acceleration both at the beginning and end of the maneuver, $t=0$ and $t=t_f$, and is given by~\cite{Caverly2014Traj}
\begin{equation}
\label{eq:VaryingTdes}
T_{des}(t)=\left[10 \left( \frac{t}{t_f}\right)^3-15 \left( \frac{t}{t_f}\right)^4 + 6 \left( \frac{t}{t_f}\right)^5\right]\left(T_f-T_0\right)+T_0,
\end{equation}
where $t_f$ is the duration of the desired maneuver, $T_f$ is the equilibrium tension associated with the desired tip displacement at the final time $t_f$, and $T_0$ is the equilibrium tension that corresponds to the initial tip displacement at $t = 0$ based on Figure~\ref{fig:desired}. The effect that this smooth time-varying feedforward input has compared to a simpler constant feedforward input based on the final desired tip deflection (i.e., $T_{des}(t) = T_f$) is assessed in the simulation results of Section~\ref{sec:simulation}.

\section{Simulation Results} \label{sec:simulation}

In this section, simulation results of the proposed PD controller with a constant feedforward input and a time-varying feedforward input are compared.
Also, the effect of constraining the control input to remain a non-negative tension throughout the simulation is investigated.

The numerical parameters from Table~\ref{table:SimParam} are used in the simulation of the system's nonlinear dynamics.
All simulations are run for $200$~seconds, starting with the initial tip deflection of $1$~m. The desired setpoint for the tip deflection is set to $w_{des}(L)=1.26855$~m, corresponding to the desired tension $T_{f}=1$~N, chosen based on the relationship in Figure~\ref{fig:desired}.

\subsection{PD Control with Constant Feedforward Input}
\label{sec:SimPDConstant}
\begin{figure}[!t]%[htb]
	\centering
	\subfigure[]
	{
		\includegraphics[width=0.48\textwidth]{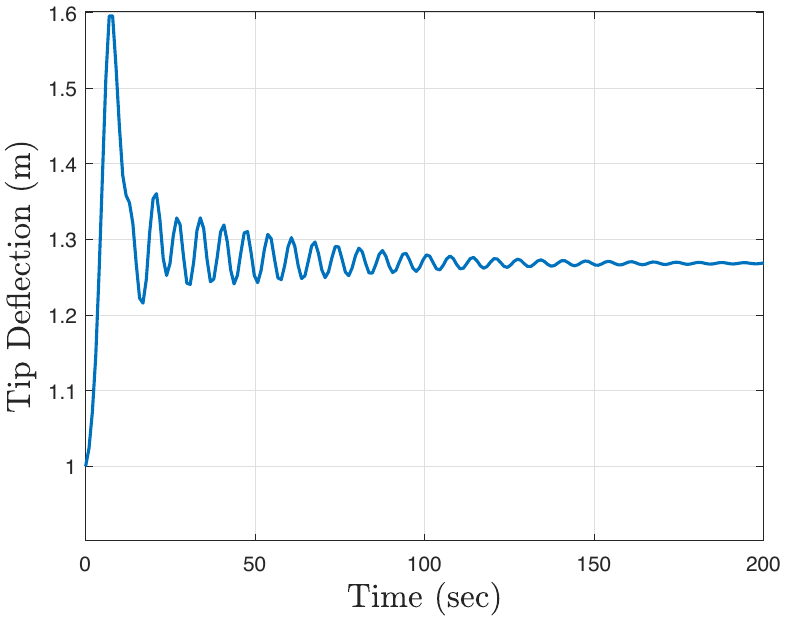}
		\label{subfig:Non1N}
	}
	\subfigure[]
	{%[width=3.5in]
		\includegraphics[width=0.48\textwidth]{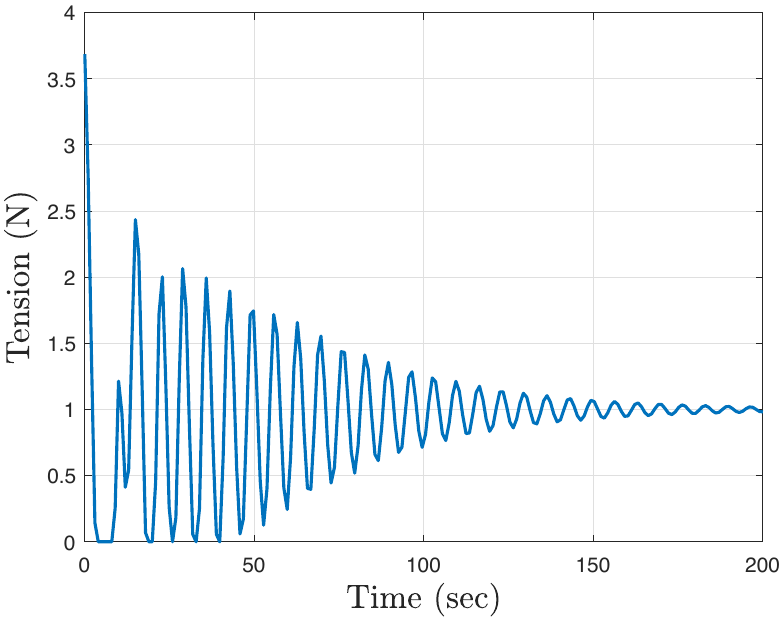}
	}
	\subfigure[]
	{
		\includegraphics[width=0.48\textwidth]{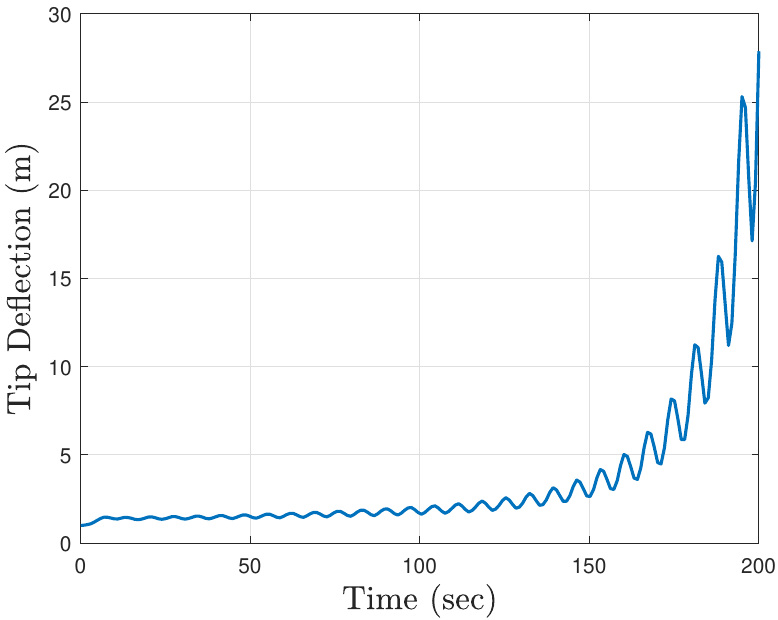}
		\label{subfig:Non1NUnstable}
	}
	\subfigure[]
	{%[width=3.5in]
		\includegraphics[width=0.48\textwidth]{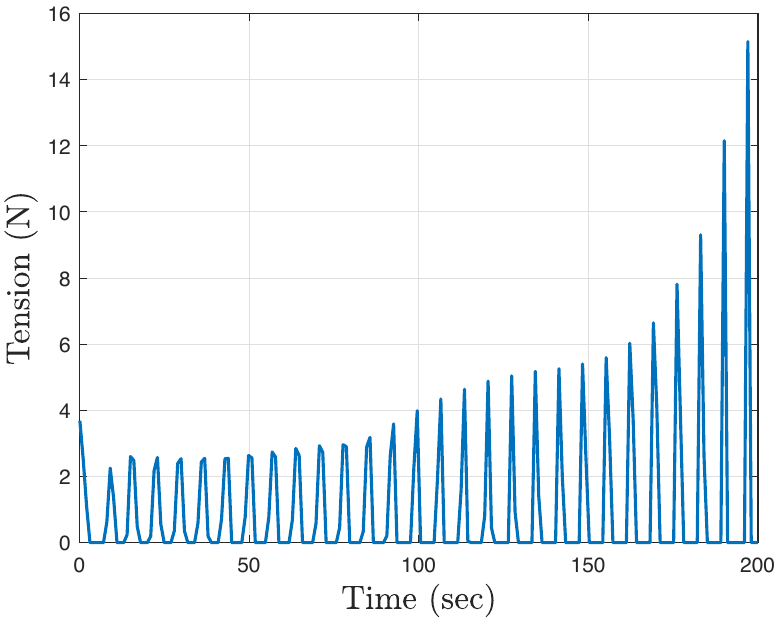}
	}
	\caption{Simulation results of PD control with the constant feedforward input. The proportional control gain is $k_p=10$~N/m and the derivative control gain is $k_d=25$~N$\cdot$s/m for (a) and (b), and $k_d=50$~N$\cdot$s/m for (c) and (d). The tip deflections versus time are shown in (a) and (c), while the control input versus time are shown in (b) and (d).}
    \label{fig:Non1N}
\end{figure}

Simulations of the proposed PD controller with the constant feedforward input $T_{des}(t) = 1$~N and constant desired tip deflection $w_{des}(L) = 1.26855$~m are performed first with two different values of derivative control gains to determine the effect the choice of control gains has on the stability and performance of the closed-loop system.
Both cases use the same proportional control gain of $k_p = 10$~N/m.
The simulation result with a smaller derivative control gain of $k_d = 25$~N$\cdot$s/m
 in Figure~\ref{subfig:Non1N} shows that the PD controller is effective in damping out the oscillations caused by the initial step input in desired tip deflection.
 However, the nonlinear system becomes unstable with larger derivative control gain $k_d = 50$~N$\cdot$s/m, as shown in Figure~\ref{subfig:Non1NUnstable}. 

 The unstable closed-loop response with the larger derivative control gain appears to contradict the guarantee of robust closed-loop input-output stability provided by the passivity theorem. However, it is important to note that the passivity analysis performed in Section~\ref{sec:PassivityAnalysis} only considers the linearized system dynamics, whereas the simulations are performed on the nonlinear system dynamics. This indicates a likelihood that implementing a step input in desired tip deflection and feedforward input results in significant perturbations from the desired equilibrium point where the linearized dynamics no longer provide a sufficient description of the system's dynamics. The nonlinear system dynamics are likely no longer passive in this case and all closed-loop stability guarantees no longer stand. This motivates the need for a time-varying feedforward input and desired tip deflection.

\subsection{PD Control with Time-Varying Feedforward Input}

\begin{figure}[!t]
	\centering
	\subfigure[]
	{
		\includegraphics[width=0.48\textwidth]{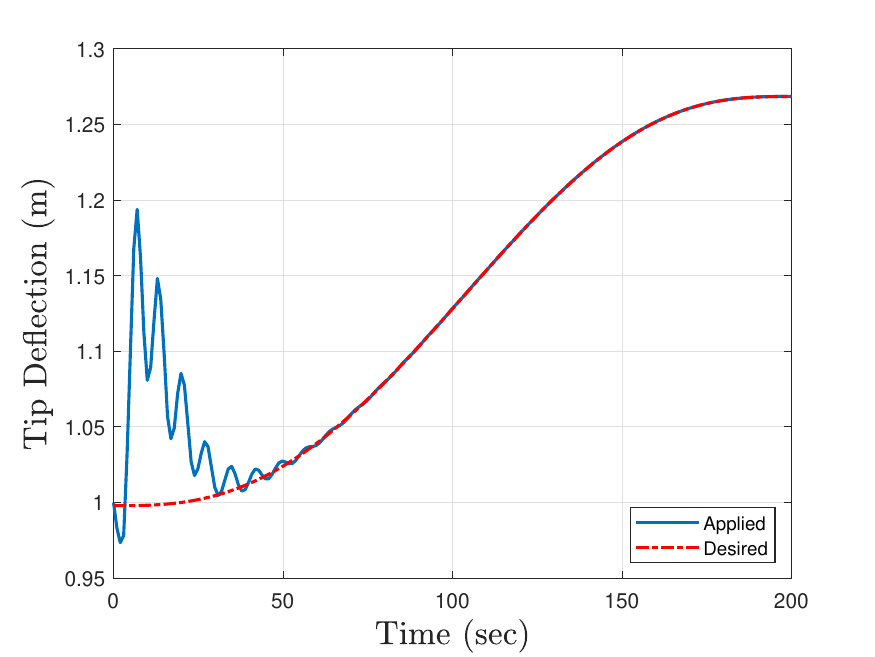}
	}
	\subfigure[]
	{
		\includegraphics[width=0.48\textwidth]{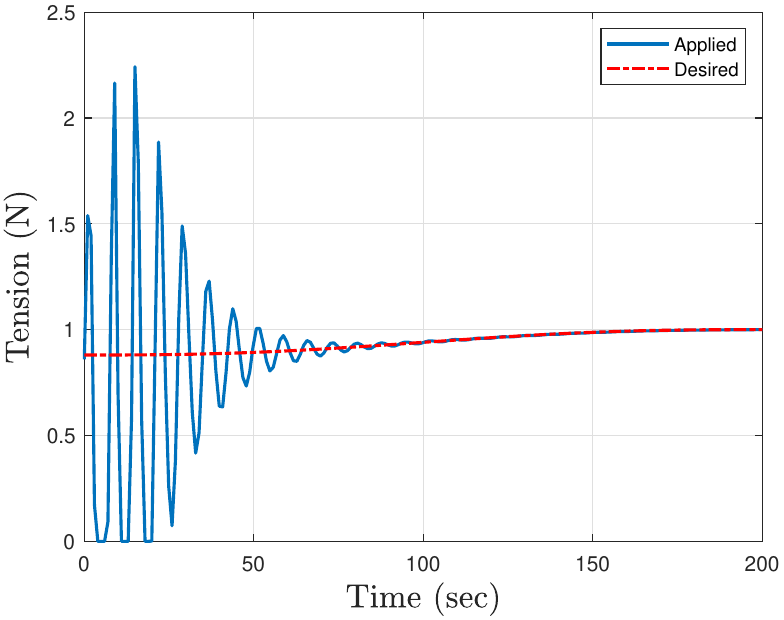}
	}
	\caption{Simulation results of PD control with the time-varying feedforward input using $k_p=10$~N/m and $k_d=50$~N$\cdot$s/m with plots of (a) tip deflection and (b) control input versus time.}
	\label{fig:Non1NVarying}
\end{figure}

The time-varying feedforward input described in~\eqref{eq:VaryingTdes} is implemented with the proposed PD controller and a time-varying desired tip deflection of
\begin{equation}
\label{eq:VaryingWdes1}
w_{des}(L,t)=\left[10 \left( \frac{t}{t_f}\right)^3-15 \left( \frac{t}{t_f}\right)^4 + 6 \left( \frac{t}{t_f}\right)^5\right]\left(w_{f}-w_{0}\right)+w_{0},
\end{equation}
where $w_{0} = 1$~m is the initial boom tip deflection at the initial time and $w_{f} = 1.26855$~m is the desired boom tip deflection at the final time. The same PD control gains used in Section~\ref{sec:SimPDConstant} are implemented in this simulation, where the nonlinear closed-loop system does not become unstable and converges to the desired setpoint in the desired time, as shown in Figure~\ref{fig:Non1NVarying}.
It was also demonstrated through further tuning that the closed-loop system remained stable for all reasonable choices of control gains, which demonstrates that the time-varying feedforward recovers the robustness provided by the proposed passivity-based PD control strategy~\cite{lee2024robust}. This is most likely due to the fact that the time-varying feedforward input and desired tip deflection allow for the closed-loop dynamics to remain close at all time to their linearized representation, which is known to be passive.

\section{Experimental Test Results} \label{sec:test}

In this section, the proposed PD controller is implemented and validated on a small-scale prototype of the CABLESSail system. This section proceeds with a description of the prototype, followed by results comparing the performance of the proposed controller with and without the use of an experimentally-determined mapping from motor torque to boom tip deflection.

\subsection{CABLESSail TRAC Boom Prototype Testbed}
\begin{figure}[t!]
    \centering
	\subfigure[]%[height=6.5cm]
	{    \includegraphics[width=.8\linewidth]{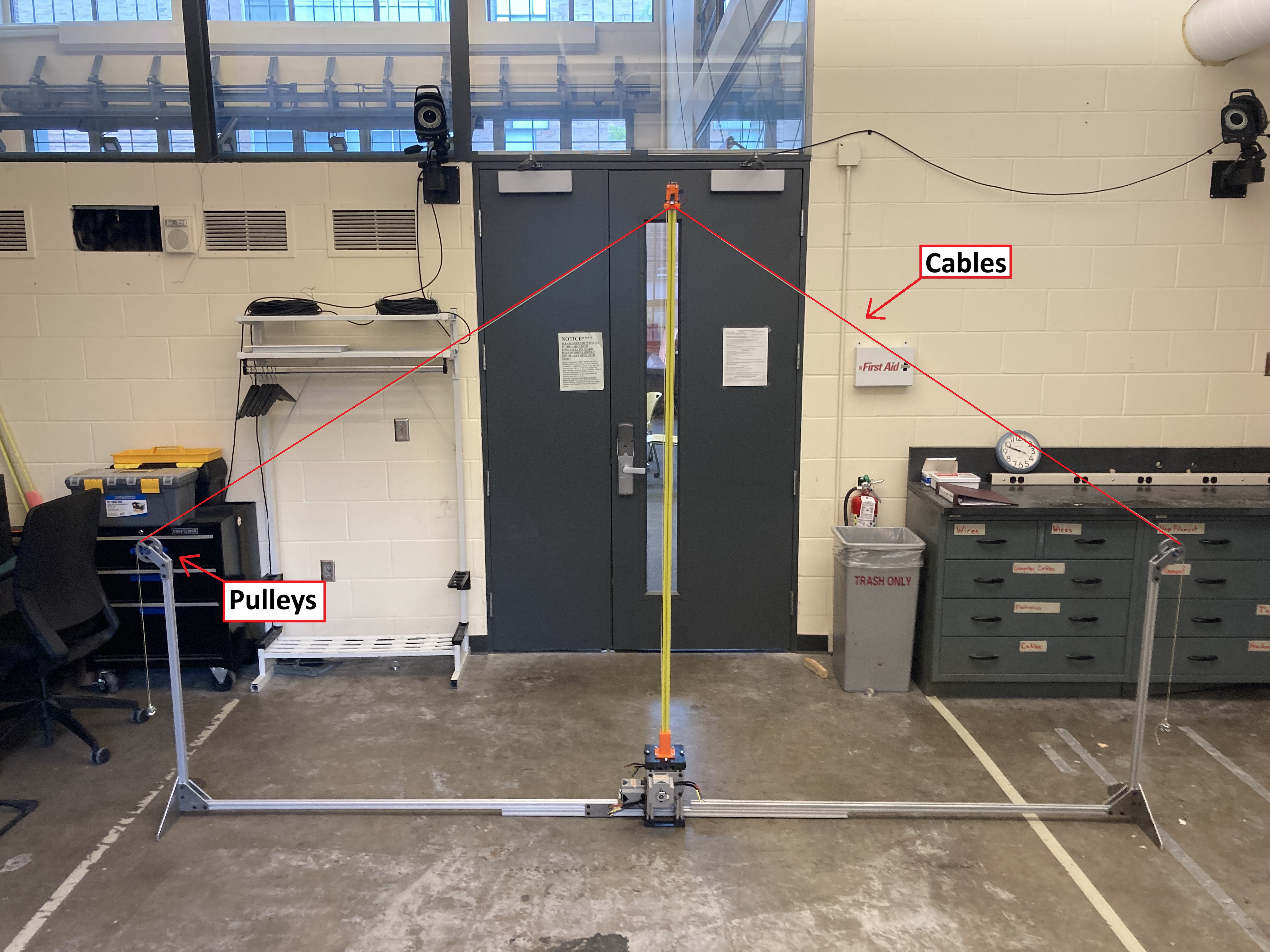}
        \label{subfig:SailTension}
    }
    \subfigure[]
    {
        \includegraphics[width=.8\linewidth]{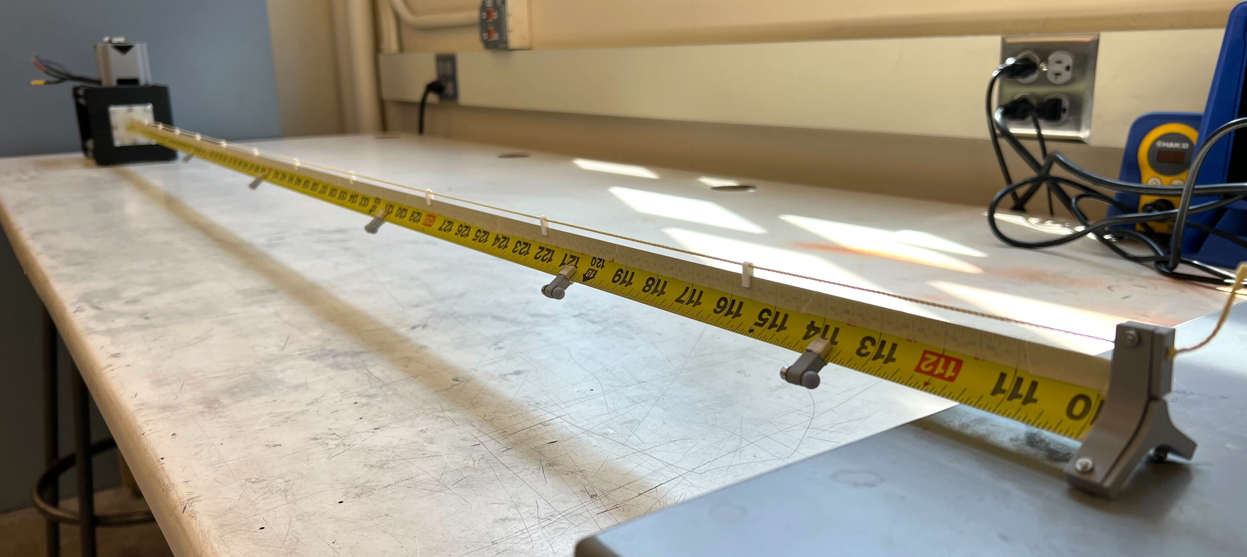}
        \label{subfig:tracboomtapemeasure}
    }
    \caption{Images of (a) the CABLESSail prototype testbed with the Vicon motion capture system in the background and (b) a close-up of the TRAC boom prototype with a single actuating cable. The spreaders on the photo were removed before the experimental tests.}
    \label{fig:testbed}
\end{figure}

To validate the proposed control law, the small-scale CABLESSail prototype developed in~\cite{Bodin2024SciTech} is used for experimental tests. The testbed shown in Figure~\ref{fig:testbed} is composed of a single TRAC boom, a cable along the web of the boom, an actuating motor, an IMU module, two cables at the side of the prototype, and supporting electronics. The single boom is fabricated using two tape measures to imitate a TRAC boom of Solar Cruiser with a scaled-down length of $1.61$~m. The cable is attached to the end of the boom and connected to the motor at the bottom, which gives tension to the cable. No spreaders are included on the prototype, which allows the cable to connect directly from the actuating winch to the boom tip. The IMU module is mounted on the tip of the boom to measure the tip deflection using a nine-degree-of-freedom Adafruit BNO-055 equipped with a rate gyro. Two additional cables are attached at the end of the boom and are maintained at roughly constant tensions through the use of hanging masses to emulate the forces applied to the boom by the solar sail's membrane.
Ground-truth boom tip deflection data is obtained by a Vicon motion capture system. Further details on the design of the prototype can be found in~\cite{Bodin2024SciTech}. 

In order to estimate the tip deflection in real time for use with the proposed feedback controller, the colored noise Kalman filter developed in~\cite{Bodin2024SciTech} is used. This Kalman filter uses the motor encoder and the tip-mounted IMU along with a model of the frequency spectrum of the motor encoder noise.

\subsection{Experimental Test Conditions}
The experimental tests are designed to provide insight to the use of CABLESSail on a full-scale Solar Cruiser solar sail. To this end, a desired boom tip deflection of $6$~mm is chosen, which corresponds to roughly a $11$~cm tip deflection in a full-scale $29.4$~m long Solar Cruiser boom following the scaling laws outlined in~\cite{Bodin2024SciTech}. It was shown in~\cite{gauvain2023solar} that the largest expected boom tip deflection that would need to be canceled out on Solar Cruiser to negate unwanted disturbance torques is $10$~cm, which justifies the choice of desired tip deflection for these experiments.

The equilibrium motor torque required to hold the boom tip at a deflection of $6$~mm is found to be $0.327$~N$\cdot$m through open-loop experimental trials. Through similar experimental tests it is determined that the boom tip does not deflect when torques of $0.15$~N$\cdot$m or less are applied, which defines the minimum static torque to be applied during the experiments. When implementing the constant feedforward input during tests, the feedforward torque is set to $T_{des}(t) = 0.327$~N$\cdot$m. When implementing the time-varying feedforward input, the expression in~\eqref{eq:VaryingTdes} is substituted with the initial and final values of $T_0=0.15$~N$\cdot$m and $T_f=0.327$~N$\cdot$m, respectively.

\subsection{PD Control with Constant and Time-Varying Feedforward}
\begin{figure}[!t]%[htb]
	\centering
	\subfigure[]
	{%[width=3.5in]
		\includegraphics[width=0.7\textwidth]{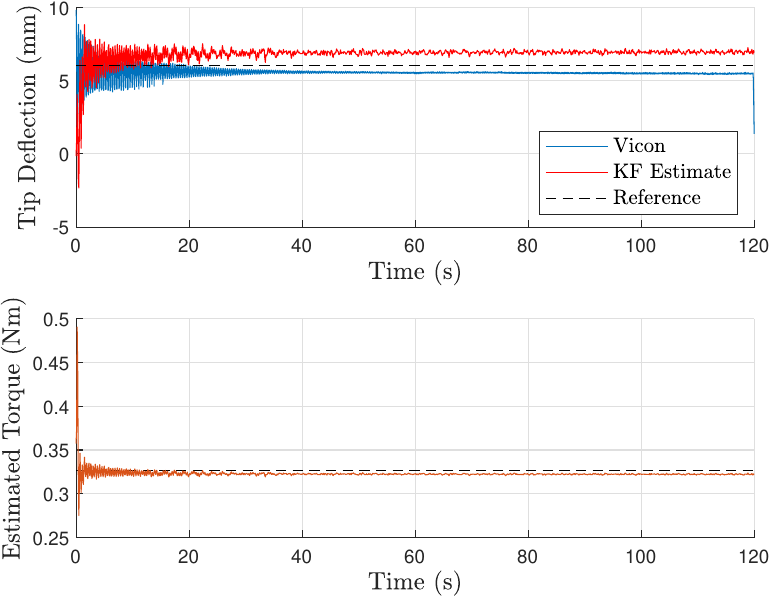}
		\label{subfig:PD_constFF}
	}
	\subfigure[]
	{
		\includegraphics[width=0.7\textwidth]{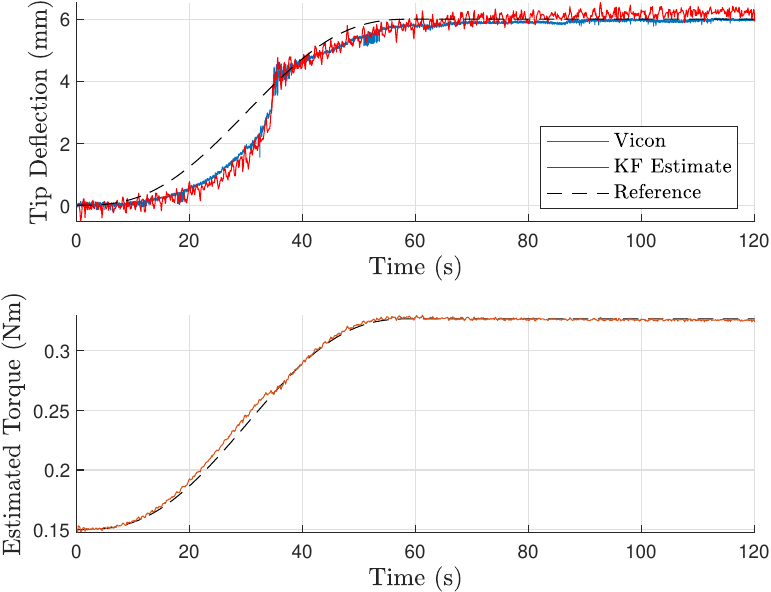}
		\label{subfig:PD_TVFF_woMap}
	}
	\caption{Experimental result of the PD control with (a) constant and (b) time-varying feedforward inputs, respectively. The control gains are $k_p=0.005$~N$\cdot$m/mm and $k_d=0.01$~N$\cdot$s$\cdot$m/mm, and the mapping between the torque and the tip deflection is not included in this test.}
	\label{fig:Test_woMap}
\end{figure}

The first experimental test involves the use of a constant feedforward input with a constant desired boom tip deflection of $6$~mm. The second test makes use of a time-varying feedforward input and a time-varying desired boom tip deflection defined by~\eqref{eq:VaryingWdes1} with $w_{0} = 0$~mm and $w_{f} = 6$~mm. PD control gains of $k_p=0.005$~N$\cdot$m/mm and $k_d=0.01$~N$\cdot$s$\cdot$m/mm are used in both tests. Plots of the boom tip deflection and estimated torque applied by the motor during the test are provided in Figure~\ref{fig:Test_woMap}.

Based on Figure~\ref{fig:Test_woMap} it is observed that the PD control with both feedforward input options works well, even in the experimental setting with significant model uncertainty. However, the comparison between the two feedforward options emphasizes the advantage of the time-varying feedforward input over the constant feedforward input. PD control with the constant references in Figure~\ref{subfig:PD_constFF} requires a much larger control torque at the beginning of the test and takes more time to settle down from the initial oscillation, while PD control with the time-varying feedforward input and the time-varying desired boom tip deflection in Figure~\ref{subfig:PD_TVFF_woMap} exhibits significantly less vibrations during the test. However, it is also observed in Figure~\ref{subfig:PD_TVFF_woMap} that closed-loop response of system significantly lags the time-varying desired boom tip deflection. This motivates further investigation into the mapping from the feedforward input torque to the desired boom tip deflection.

\subsection{PD Control with Improved Desired Boom Tip Deflection Mapping}
\begin{figure}[!t]%[htb]
	\centering\includegraphics[width=3.5in]{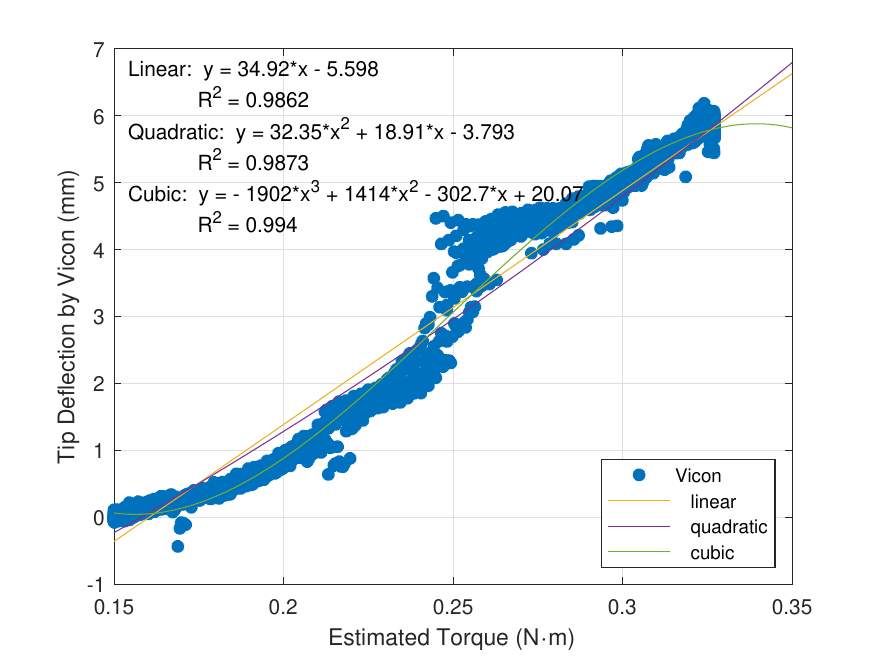}
	\caption{Mapping of the torque to the tip deflection based on three open-loop tests.}
	\label{fig:mapping_test}
\end{figure}

The lag observed in Figure~\ref{subfig:PD_TVFF_woMap} points to the importance of obtaining an accurate mapping between the torque and the tip deflection. In other words, it is likely that the time-varying feedforward torque input applied is not consistent with the time-varying desired boom tip deflection during the transient portion of the trajectory. Unlike the simulation where an analytic model with known properties is available, it is difficult to obtain a mapping like the one shown in Figure~\ref{fig:desired} from an analytic dynamic model. In order to better model the relationship between motor torque and boom tip deflection, multiple open-loop tests are performed where the motor torque applied is increased incrementally with time. The data collected from these test are shown in Figure~\ref{fig:mapping_test}.

Theoretically, using linear Euler-Bernoulli beam theory, the mapping from torque applied to the tip deflection is expected to be quadratic, as shown in the simulation results of Figure~\ref{fig:desired}. However, because of non-ideal hardware effects, such as cogging of the brushless motors, this theoretical relationship does not necessarily hold. To this end, linear, quadratic, and cubic relationships are fit to the data in Figure~\ref{fig:mapping_test}. The cubic relationship appears to be the best at capturing the nonlinear effects present in the hardware and is therefore used for subsequent testing. Based on this finding, the desired tip deflection is calculated by mapping the time-varying feedforward input torque through the cubic fit found in Figure~\ref{fig:mapping_test}, resulting in
\begin{multline}
\label{eq:VaryingWdes}
w_{des}(L,t) = -(1902~\text{mm}/(\text{N}\cdot\text{m})^3) \left(T_{des}(t)\right)^3+(1414~\text{mm}/(\text{N}\cdot\text{m})^2)  \left(T_{des}(t)\right)^2 \\ -(302.7~\text{mm}/(\text{N}\cdot\text{m}))  T_{des}(t)+20.07~\text{mm},
\end{multline}
where $T_{des}(t)$ is obtained from~\eqref{eq:VaryingTdes}. Since $T_{des}(t)$ is a fifth-order polynomial,~\eqref{eq:VaryingWdes} results in a fifteenth-order polynomial function of time. This is significantly higher order than the fifth-order polynomial for the time-varying desired tip deflection in~\eqref{eq:VaryingWdes1}, which noticeably increases the slope of the desired tip deflection between its initial and final values.

\begin{figure}[!t]%[htb]
	\centering\includegraphics[width=0.7\textwidth]{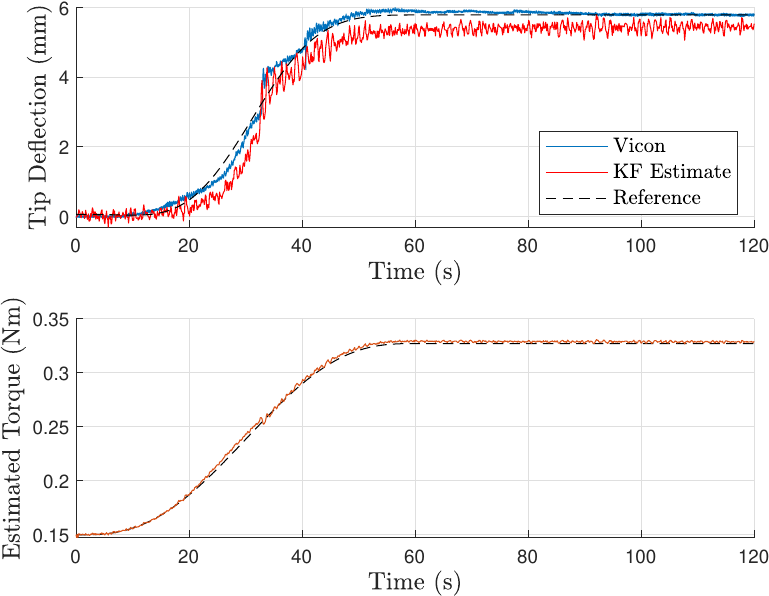}
	\caption{Experimental result of PD control with the time-varying feedforward input. The control gains are $k_p=0.005$~N$\cdot$m/mm and $k_d=0.01$~N$\cdot$s$\cdot$m/mm, and the cubic mapping of the torque to the tip deflection in Figure~\ref{fig:mapping_test} is used to calculate the desired tip deflection.}
	\label{fig:PD_TVFF_wMap}
\end{figure}

Applying the new time-varying desired boom tip deflection of~\eqref{eq:VaryingWdes}, the proposed PD controller with the time-varying feedforward input is implemented experimentally, with results shown in Figure~\ref{fig:PD_TVFF_wMap}. The tracking performance is significantly improved compared to the closed-loop performance in Figure~\ref{subfig:PD_TVFF_woMap} without the experimentally-determined mapping from torque to boom tip deflection. 

\subsection{Discussion}
Similar to the simulation results in Section~\ref{sec:simulation}, the experimental tests show that PD control with the time-varying feedforward input works well, both to achieve the desired tip deflection and to minimize unwanted vibrations in the boom.
The experimental results are particularly promising, considering the large amount of uncertainty in the boom tip deflection estimate and the system dynamics due to the lack of an accurate analytical model and the use of low-cost equipment.

Interestingly, the performance of the proposed controllers are noticeably better in the experimental results of Figures~\ref{subfig:PD_TVFF_woMap} and~\ref{fig:PD_TVFF_wMap} compared to the simulation results of Figures~\ref{subfig:Non1N} and~\ref{fig:Non1NVarying}. In particular, far fewer vibrations are induced in the experiments and convergence to the desired boom tip deflection occurs quicker. This is likely due to differences in material properties, as well as the presence of air resistance and natural damping in the experimental prototype. The simulated effect of sail tension is also present in the experimental test setup, but not in the simulation.

As seen in Figures~\ref{subfig:PD_TVFF_woMap} and~\ref{fig:PD_TVFF_wMap}, the estimate of the boom tip deflection obtained by the Kalman filter from~\cite{Bodin2024SciTech} has inaccuracies compared to the ground-truth Vicon data. For the purposes of this paper that focuses on the design and analysis of a feedback controller, this is not a significant issue and, in fact, suggests that the proposed PD controller performs well in the presence of measurement noise and uncertainty. However, improvements to the state estimation are certainly possible and will be explored in future work on this topic.

There are hardware limitations of the experimental testbed that affect the closed-loop performance, but are not related to the control law itself. The brushless motor chosen for the tests is not designed for the small rotational motion performed in the experimental tests in this work. This results in cogging in the motor that is observed around $0.25$~N$\cdot$m in every test. It is expected that a motor with better properties would result in both a closer to ideal mapping from torque applied to boom tip deflection and smoother tracking of the desired boom tip deflection. Experiments with improved motors may be performed by the authors in future research. However, it is worth noting that the tests performed in this work provide confidence that the proposed PD controller is capable of performing the required actuation for the CABLESSail concept, even with a less-than-ideal motor.

\section{Conclusion}
This paper focused on the use of passivity-based control design to advance the proposed CABLESSail concept that uses cable-actuation to control the shape of a flexible solar sail boom. Simulation results with a Solar Cruiser boom model demonstrated the effectiveness of the proposed controller in the presence of no natural damping. Experimental tests with a small-scale prototype illustrated the ability for the controller to track a desired doom tip deflection in the presence of significant model uncertainty and actuator/sensor imperfections. A notable finding in both the simulation and experimental results is the benefit of implementing a time-varying feedforward input along with a time-varying desired boom tip deflection both in terms of closed-loop performance and stability.

Future research on this topic will determine the boom tip deflections required to generate desired control torques through an imbalance in solar radiation pressure. The PD controller developed in this work will then be implemented within this setup to reliably generate the required control torques.

\section*{Acknowledgments}
This work was supported by an Early Career Faculty grant from NASA’s Space Technology Research Grants Program under award No.~80NSSC23K0075. The authors would like to thank Keegan Bunker for providing the CABLESSail simulation code used in this work.

\bibliography{Ref_Acta}
\end{document}